\documentstyle [12pt] {article}
\newcommand{\cs}[3]{{{#3} \brace {#1 #2}}}

\newcommand{\edf}{\ {\mathop{=}\limits^{\rm def}}\ }

\begin{document}
\begin{center}
\bf {A Note on General Covariant Stability Theory }\\

\end{center}
\begin{center}
\bf{M.I.Wanas\footnote{Astronomy Department, Faculty of Science,
Cairo University, Giza, Egypt.

E-mail:wanas@frcu.eun.eg}} ; M.A. Bakry \footnote{Mathematics
Department, Faculty of Education, Ain Shams University, Cairo,
Egypt. }
\end{center}

\begin{abstract}
In the present work we suggest a general covariant theory which
can be used to study the stability of any physical system treated
geometrically. Stability conditions are connected to the magnitude
of the deviation vector. This theory is a modification of an
earlier joint work, by the same authors, concerning stability. A
comparison between the present work and the earlier one is given.
The suggested theory can be used to study the stability of
planetary orbits, astrophysical configurations and cosmological
models.

\end{abstract}
\section{Introduction}
In a previous paper [1] the authors have suggested the use of
geodesic deviation equations to study stability of gravitating
systems. In that paper, they have generalized the classical
perturbation scheme, usually used to deal with such problem. We
have suggested the use of components of the deviation vector,
representing the solution of the equation of geodesic deviation,
$$
\frac{d^2 \xi^{\alpha}}{ds^2} + 2
\cs{\beta}{\gamma}{\alpha}\frac{d \xi^{\beta}}{ds} U^\gamma
+{\cs{\beta}{\gamma}{\alpha}}_{, \lambda} U^{\beta}
U^{\gamma}\frac{d \xi^{\lambda}}{ds} = 0 \eqno{(1)}
$$
where $\xi^{\alpha}$ is the deviation vector, $U^{\beta}$ is the
unit tangent to the geodesic, $\cs{\beta}{\gamma}{\alpha}$ is the
Christoffel symbol of the second kind and $(s)$ is an invariant
parameter. Now, $\xi^{\alpha} (s)$ is the solution of equation (1)
in the interval $[a,b]$ in which the functions $\xi^{\alpha} (s)$
behave monotonically. This vector
  reflects the reaction of
the system under perturbation.  The quantities, that have been
suggested in [1], to be used as  sensors for stability of the
system, are
$$
q^{\alpha}\edf \lim_{s\rightarrow b}\xi^{\alpha}(t).  \eqno{(2)}
$$
 The criterion suggested is that,  if $q^\alpha \rightarrow \infty$,
the system would be unstable, otherwise it would be stable. This
criteria has been used to study stability of a number of
cosmological models. Applications in cosmological models, using
this criterion, is somewhat easy since most of these models depend
on one function, the scale factor. Further applications show the
non-covariance, of the scheme, under coordinate transformations.

It appears that if stability conditions are obtained depending on
the quantities (2), these conditions would not be, in general,
covariant . This is because the components of the deviation vector
depend on the coordinates system used. In other words, the
stability conditions obtained would be coordinate dependent. We
are going to call the scheme suggested in [1] the {\it "Coordinate
Dependent Scheme"}, (CDS).

The aim of the present note is to modify the quantity (2) in order
to get covariant stability conditions.
\section{Covariant Stability Conditions}
To get covariant results, independent of the coordinate system
used, one has to replace the contravariant components of the
deviation vector used in (2) by its magnitude, then we examine the
limit
$$
q \edf \lim_{s\rightarrow
b}(\xi^{\alpha}\xi_{\alpha})^{\frac{1}{2}}. \eqno{(3)}
$$
Now, if $q \rightarrow \infty$, then the
system is unstable. Otherwise, it would be stable. \\

To summarize how to apply the covariant scheme suggested, one has
to follow the following steps:\\
1. Having a well defined problem, we solve the field equations
controlling this problem to know the type of geometry associated
with the system under consideration (the metric). \\
2. Knowing the metric of space-time, we solve the geodesic
equation to get the
unit tangent vector $U^{\alpha}$. \\
3. Using the information, obtained in the above two steps,
substituting in the geodesic
deviation equation (1) and solving it, we get the deviation vector $\xi^{\alpha}$. \\
4. Evaluating the scalar $\xi^{\mu}\xi_{\mu}$ and examining its
limit as given by (3), one can answer stability question. \\ If
$q\rightarrow \infty$, the system will be unstable. Otherwise,
it will be stable. \\
5. A strong stability condition can be achieved if,
$$
\lim_{t\rightarrow \infty}(\xi^{\alpha}
\xi_{\alpha})^{\frac{1}{2}} =0. \eqno{(3)}
$$
We are going to call this scheme {\it "The Coordinate Independent
Scheme"}, (CIS).
\section{Discussion}
If we use the scheme  suggested in the present work CIS and apply
it to some of the world models examined in the previous work [1]
we get the results that are summarized and compared, to those
obtained using the CDS, in Table 1. In this table,  the
cosmological models treated are classified as follows. The first
set of models  represents world models constructed using "General
Relativity" (GR). In the second set, we examine a world model
depending on "Miln Kinematical Relativity" (KR) and another one
constructed using "Brans-Dicke Theory" (BD). The third set
contains models resulting from "M$\o$ller's Tetrad Theory of
Gravitation" (MTT)[2]. The last set contains models obtained using
the "Generalized Field Theory" (GFT) [3]. The sample, in Table 1,
is chosen in such a way that it represents models depending on
different geometric field theories. It is clear from the following
table that the use of the covariant scheme, suggested in the
present work, gives results, some of which are different from
those obtained in the previous work.
\newpage
\begin{center}
Table1:  Stability of Some World Models  Using CDS and CIS. \\
\begin{tabular}{|c|c|c|c|}  \hline \hline
&&& \\
Theory & Model  & CDS  & CIS\\
&&& \\
\hline \hline
& & & \\
GR & Einstein [4] & Unstable & Unstable\\
 & De Sitter [4] & Stable & Unstable\\
 & Einstein-De Sitter [4]& Stable& Unstable \\
 & Radiation [5]& Unstable & Unstable\\
& & & \\
 \hline
 &&& \\
KR & Miln [4]& Stable & Stable\\
 BD& Brans-Dick [6]& Stable & Unstable\\
&&& \\
 \hline
 & & & \\
MTT &$D < 0$ [7]& Stable& Unstable \\
& $D>0$ [7]& Conditional& Unstable \\
&&& \\
 \hline
 & & & \\
GFT  & $k=-1$ [8]& Stable & Stable \\
 & $k=0$ [8] & Unstable & Unstable \\
&&& \\
 \hline
\end{tabular}
\end{center}
  The similar results
obtained, using the suggested scheme and the previous one [1], are
just coincidence. It is obvious that changing the coordinate
system used to construct a world model will not affect the results
of the last column of Table 1, while it may change those given in
the third column.

The scheme suggested in the present work has been successfully
used to study stability of non-singular black holes [9]. Further
details will be published elsewhere.

\section*{References}
{[1]} Wanas, M.I. and Bakry, M.A. (1995) Astrophys. Space Sci.
{\bf{228}}, 239. \\
{[2]} M$\o$ller, C.(1978) Mat. Fys. Skr. Dan. Vid. selk.
{\bf{39}},13, 1. \\
{[3]} Mikhail, F.I. and Wanas, M.I. (1977) Proc. Roy. Soc. Lond. A
{\bf{356}}, 471. \\
{[4]} McVittie, G.C. (1961) "{\it{Facts and Theory of
Cosmology}}", Eyre \& spittswoode, London. \\
{[5]} Sciama, D.W. (1971) "{\it{Modern Cosmology}}", Cambridge, London. \\
{[6]} Wienberg, S. (1972)"{\it{Gravitation and Cosmology}}" John Wily \& Sons. \\
{[7]} Saez, D. and de-Juan , T. (1984) Gen.Rel Grav. {\bf{16}}, 5.
\\
{[8]} Wanas, M.I. (1989) Astrophys. Space Sci. {\bf{154}}, 165. \\
{[9]} Nashed, G.G.L. (2003) Chaos, Solitons and Fractals, {\bf
{15}}, 841. \\
\end{document}